\documentclass[12pt, a4paper]{article}
\usepackage[utf8]{inputenc}
\usepackage{physics} 
\usepackage{amsmath}
\usepackage{amsfonts}
\usepackage{amssymb}
\usepackage{graphicx}
\usepackage{authblk}
\usepackage{xcolor}
\usepackage{subcaption}
\usepackage{cite}
\usepackage{multirow}
\usepackage{orcidlink}
\DeclareUnicodeCharacter{0327}{\c{}}
\PassOptionsToPackage{colorlinks=true,linkcolor=blue,citecolor=blue,urlcolor=blue}{hyperref}

\usepackage[left = 1.5cm, right = 1.5cm, top = 1cm, bottom = 2cm]{geometry}
\begin{document}

\begin{center}
{}

{\large{\bf Cosmological Perturbations and Observational Constraints on Spinor Field Quintessence Dark Energy}}
\vskip .5cm
\textbf{ Mahendra Goray \orcidlink{0000-0002-3529-7030} $^{a}$}

\vskip0.1cm

$^a$Department of Physics, S.P. College, Sido Kanhu Murmu University, Dumka-814101, Jharkhand, India\\

\textbf{E-mail:} goraymahendra92@gmail.com  \\
\end{center}

\begin{abstract}
We investigate the linear perturbation cosmology of a spinor-field realization of quintessence dark energy by implementing the model in the Einstein--Boltzmann solver \texttt{CLASS}. The model parameters are constrained using a Markov Chain Monte Carlo analysis with Pantheon$^{+}$ Type Ia supernovae, cosmic chronometers, DESI DR2 baryon acoustic oscillations, redshift-space distortions, and Planck 2018 CMB distance-prior data. For the full combined dataset, we obtain $w_{\rm de}=-0.9710^{+0.0206}_{-0.0206}$, $\Omega_{m0}=0.2939^{+0.0047}_{-0.0047}$, and $H_0=66.23^{+0.55}_{-0.53},\mathrm{kms^{-1}Mpc^{-1}}$. The corresponding perturbation quantities are $\sigma_8=0.7537^{+0.0076}_{-0.0075}$ and $S_8=0.7459^{+0.0067}_{-0.0065}$. We find that the spinor-quintessence model closely reproduces the predictions of a phenomenological constant-$w$ model in the matter power spectrum, growth-rate observable $f\sigma_8$, growth index, and CMB temperature anisotropy spectrum, with deviations generally below the percent level. A comparison with $\Lambda$CDM and $w$CDM shows statistically indistinguishable fits, although the Bayesian information criterion favors the simpler $\Lambda$CDM model. Our results demonstrate that the spinor-field quintessence scenario remains consistent with current expansion-history and structure-growth observations when its perturbation evolution is treated consistently, providing a field-theoretically motivated realization of constant-$w$ dark energy at the background and linear perturbation levels.
\\
\\
\textbf{Keywords:} \texttt{CLASS}; Dark energy; Spinor field; Perturbation; Quintessence; Parametrization. 
\end{abstract}

\section{Introduction}

The discovery of the late-time accelerated expansion of the Universe has established one of the central challenges of modern cosmology \cite{Riess1998, Perlmutter1999}. Observations of Type Ia supernovae \cite{Pantheon2018, PantheonPlus2022}, together with measurements of the cosmic microwave background (CMB) \cite{Planck2013, Planck2018}, baryon acoustic oscillations (BAO)\cite{eBOSS2021, DESI2024, DESI2023}, large-scale structure (LSS)\cite{Eisenstein2005, Alam2017, AG2024}, and gravitational lensing\cite{KiDS2021, DES2022}, indicate that the Universe is currently dominated by a component with negative effective pressure, commonly referred to as dark energy. Within the standard cosmological framework, dark energy accounts for approximately $70\%$ of the present-day energy budget, while dark matter contributes most of the remaining non-relativistic matter component \cite{Planck2018}. Despite their dominant contribution to the cosmic energy density, the fundamental nature of both dark energy and dark matter remains unknown.

The simplest and most successful description of the observed cosmic expansion is provided by the $\Lambda$CDM model, in which dark energy is represented by a cosmological constant with equation of state (EoS) $w\equiv p/\rho=-1$ \cite{Weinberg1989, Carroll1992, Carroll2001, Copeland2006}. Although $\Lambda$CDM provides an excellent fit to a wide range of cosmological observations, its physical interpretation is accompanied by longstanding theoretical difficulties, including the cosmological constant and coincidence problems \cite{Carroll1992, Riess2022, DiValentino2021, Verde2019}. These issues have motivated extensive investigations of dynamical dark-energy scenarios in which the effective EoS can evolve with cosmic time. Among the simplest extensions is the canonical quintessence model, in which a scalar field evolves under a potential and generally produces $w\neq -1$ \cite{Ratra1988, Wetterich1988, Caldwell1998}. At the phenomenological level, a widely used parametrization is the Chevallier--Polarski--Linder (CPL) form, $w(a)=w_0+w_a(1-a)$, which provides a convenient two-parameter description of a time-dependent dark-energy EoS \cite{Chevallier2001, Linder2003}.

Recent observations have further strengthened the motivation for exploring departures from the minimal $\Lambda$CDM scenario. In particular, persistent discrepancies in the inferred value of the Hubble constant $H_0$\cite{Riess2022, Verde2019}, and in the amplitude of matter clustering, commonly characterized by $S_8$\cite{Asgari2021, Abbott2022, Amon2022}, have stimulated considerable interest in extensions of the standard cosmological model. Moreover, recent DESI BAO measurements \cite{DESIDR1, DESIDR2}, together with CMB and supernova observations, provide increasingly precise constraints on the expansion history and the growth of cosmic structure\cite{PantheonPlus2022, Planck2018, AG2024, DES2022}. These developments emphasize the importance of testing dark-energy models not only at the homogeneous background level but also through their predictions for cosmological perturbations and structure formation.

Most phenomenological dark-energy models are formulated by specifying an effective energy density and pressure, or equivalently an EoS, within general relativity. Although such descriptions are useful for confronting observations, they do not necessarily identify the underlying microscopic field responsible for the dark-energy sector\cite{Carroll2001, Copeland2006, Ratra1988, Wetterich1988, Caldwell1998, Chevallier2001, Linder2003}. A field-theoretic description therefore provides a complementary framework in which the origin of the effective energy density, pressure, and perturbations can be studied consistently. In this context, spinor fields offer an interesting alternative to the more extensively studied scalar-field realizations of dark energy. A spinor field minimally coupled to gravity contributes to the energy-momentum tensor and can generate an effective dark-energy component through its dynamics governed by the Dirac equation. Consequently, spinor-field cosmology provides a direct field-theoretic realization of dynamical dark energy within the framework of general relativity \cite{Ribas2005, Kremer2006, Saha2018, SahaPRD2001, bsaha2025}.

In our previous studies, we investigated spinor-field realizations of several dark-energy scenarios, including the generalized Chaplygin gas (GCG) \cite{goray2025}, modified Chaplygin gas (MCG) \cite{goray2026}, and quintessence-type models \cite{goray2026NPB}, and confronted their background cosmological predictions with observational data. These analyses demonstrated that spinor-field models can reproduce viable late-time expansion histories and provide competitive constraints on their phenomenological parameters. However, agreement at the background level alone is insufficient to establish the cosmological viability of a dark-energy model. The evolution of density and velocity perturbations determines the growth of structure, matter clustering, gravitational potentials, and the formation of secondary anisotropies in the CMB. Therefore, a consistent assessment of a field-theoretic dark-energy model requires its perturbative dynamics to be incorporated into a Boltzmann framework and confronted with observations sensitive to structure growth.

In this work, we extend our previous background-level analysis by considering the perturbation dynamics of the spinor field quintessence-type dark energy (SPINOR QUINT) model \cite{goray2026NPB}. We implement the model in the \texttt{CLASS} (Cosmic Linear Anisotropy Solving System) Einstein--Boltzmann solver, thereby allowing us to follow its effects on the evolution of cosmological perturbations and the resulting observables \cite{Blas2011}. In particular, we investigate the impact of the SPINOR QUINT field on the matter power spectrum, the growth of large-scale structure, and the CMB anisotropies. We confront the model with a combination of recent and complementary cosmological observations, including Pantheon$^+$ Type Ia supernovae (PP), cosmic chronometers (CC), baryon acoustic oscillations (BAO), redshift-space distortion (RSD) measurements, and CMB data. The inclusion of RSD observations is especially important because they directly probe the growth of cosmic structure and therefore provide information that cannot be obtained from the homogeneous expansion history alone.

Our analysis thus provides a unified test of the SPINOR QUINT scenario at both the background and perturbation levels. By implementing the underlying spinor-field dynamics within an Einstein--Boltzmann framework and confronting its predictions with expansion, distance, CMB, and structure-growth observations, we assess whether the model can simultaneously provide a viable description of the late-time accelerated expansion and the observed growth of cosmic structure. This approach also allows us to examine whether the spinor-field realization leads to observational signatures that distinguish it from phenomenological quintessence and the standard $\Lambda$CDM framework.

This paper is organized as follows. In Section~\ref{sec:spinor}, we formulate the spinor-field realization of the constant-$w$ quintessence dark-energy model and discuss its background evolution and cosmological perturbations within the FLRW framework. The implementation of the SPINOR QUINT model in the CLASS Einstein--Boltzmann solver is also described. Section~\ref{sec:obervation} presents the observational datasets, likelihood construction, and Markov Chain Monte Carlo methodology adopted for constraining the model parameters. In Section~\ref{sec:result}, we present the observational constraints and investigate the cosmological implications of the model at both the background and perturbation levels, including the matter power spectrum, growth of structure, $\sigma_8$, $S_8$, the growth index, and CMB temperature anisotropies, with comparisons to $\Lambda$CDM and $w$CDM. Finally, Section~\ref{sec:conclusion} summarizes our main results and conclusions, and discusses possible directions for future investigations.

\section{Spinor field cosmology}
\label{sec:spinor}
\subsection{Background cosmology}
The energy density and pressure of the nonlinear spinor field function are (see References \cite{bsaha2025, goray2025, goray2026, goray2026NPB})
\begin{subequations}
\begin{align}
\rho_{\rm de}&=\lambda F,\\
p_{\rm de}&=\lambda\left(2KF_K-F\right),
\end{align}
\end{subequations}
where, $\lambda$ denote the the self-coupling constant, and $F_K=dF/dK$. The nonlinear spinor contribution $F(K)$ which are generated from real bilinear forms and it is associated with the Lagrangian density of the massless spinor field \cite{SahaPRD2001, goray2026, goray2026NPB}, 
\begin{eqnarray}
L_{\rm sp} = \frac{\imath}{2} \left[\bar{\psi} \gamma^{\mu} \nabla_{\mu}
\psi - \nabla_{\mu} \bar{\psi} \, \gamma^{\mu} \psi \right] - \lambda F(K). \label{lspin}
\end{eqnarray}

Assuming a constant dark-energy equation of state
\begin{equation}
p_{\rm de}=w_{\rm de}\rho_{\rm de},
\label{EOS}
\end{equation}
the nonlinear spinor interaction takes the form
\begin{equation}
F(K)=\lambda_1 K^{(1+w_{\rm de})/2},\label{F}
\end{equation}
which yields the energy density and pressure in redshift (for full derivation see Reference~\cite{goray2026}) as
\begin{subequations}
\label{mchap_eps_p}
\begin{align}
\rho_{de} &=\rho_{\rm de,0}\,(1+z)^{3(1+w_{de})}, \label{mchapsped}\\
p_{de} &= \rho_{\rm de,0}\,w_{de}\,(1+z)^{3(1+w_{de})}. \label{modchapp}
\end{align}
\end{subequations}
$\rho_{\rm de,0}=\lambda\lambda_{1}$, and $\lambda_1$ is an integration constant. Therefore, the spinor field behaves exactly as a quintessence dark-energy component with constant equation-of-state parameter $w_{\rm de}$ \cite{goray2026}.

Assuming a homogeneous and isotropic FLRW universe, the Friedmann equation becomes
\begin{equation}
H^2=\frac{8\pi G}{3}(\rho_m+\rho_{\rm de})-\frac{k}{a^2}.
\end{equation}
Introducing the present-day density parameters $\Omega_i={\rho_{i,0}}/{\rho_{\rm crit,0}}$, the normalized Hubble parameter is
\begin{equation}
E^2(z)=\Omega_{m0}(1+z)^3+\Omega_{{\rm de},0}(1+z)^{3(1+w_{\rm de})}
+\Omega_{k0}(1+z)^2,
\label{Ez}
\end{equation}
where $E(z)={H(z)}/{H_0}$. For the present analysis, we adopt a spatially flat universe, $\Omega_{k0}=0$,
so that
\begin{equation}
E^2(z)=\Omega_{m0}(1+z)^3+(1-\Omega_{m0})
(1+z)^{3(1+w_{\rm de})}.
\label{EzFlat}
\end{equation}

The above background solution constitutes the input expansion history for the perturbation calculations performed with the modified \texttt{CLASS} code.

\subsection{Cosmological perturbations}

While the background evolution determines the expansion history of the Universe, the formation of large-scale structures and the anisotropies of the cosmic microwave background (CMB) are governed by the evolution of cosmological perturbations~\cite{Planck2018, Eisenstein2005, Alam2017}. Consequently, a complete assessment of any dark-energy model requires not only agreement with background observations but also consistency with the growth of matter perturbations~\cite{Copeland2006}. To investigate the perturbative properties of the proposed SPINOR QUINT model, we implement the model in the Einstein--Boltzmann solver \texttt{CLASS}~\cite{Lesgourgues2011, Blas2011} and compute the corresponding linear cosmological observables. This subsection briefly summarizes the perturbation equations employed in the numerical analysis and describes the implementation of the model in the \texttt{CLASS} framework.

The evolution of cosmological perturbations is studied in the synchronous gauge~\cite{Ma1995}, where the perturbed Friedmann--Lemaître--Robertson--Walker metric is written as
\begin{equation}
ds^{2}=a^{2}(\tau)\left[-d\tau^{2}
+\left(\delta_{ij}+h_{ij}\right)dx^{i}dx^{j}\right],
\label{metric_sync}
\end{equation}
where $a(\tau)$ is the scale factor expressed in conformal time $\tau$, and $h_{ij}$ denotes the scalar metric perturbations. For the spinor field Quintessence model, the density contrast is defined as $\delta_{\rm sp}={\delta\rho_{\rm de}}/{\rho_{\rm de}}$, while the divergence of the peculiar velocity field is represented by $\theta_{\rm sp}$.

Assuming the spinor field behaves as an effective dark-energy fluid, the linear perturbation equations solved by \texttt{CLASS} are given by~\cite{Ma1995, Bean2004}
\begin{align}
\delta_{\rm sp}'&=-(1+w_{\rm de})\left(\theta_{\rm sp}+\frac{h'}{2}\right)
-3\mathcal{H}(c_s^2-w_{\rm de})\delta_{\rm sp},
\label{deltaeq}
\\
\theta_{\rm sp}'&=-\mathcal{H}(1-3c_s^2)\theta_{\rm sp}
+\frac{c_s^2}{1+w_{\rm de}}k^2\delta_{\rm sp},
\label{thetaeq}
\end{align}
where a prime denotes differentiation with respect to conformal time, $\mathcal{H}=a'/a$ is the conformal Hubble parameter, $k$ is the comoving wave number, and $c_s^2$ is the physical sound speed in the rest frame of the fluid~\cite{Bean2004, Hu1998}. The Einstein equations together with Eqs.~(\ref{deltaeq}) and (\ref{thetaeq}) are integrated numerically by \texttt{CLASS} to obtain the evolution of matter perturbations, transfer functions, and CMB anisotropies.

\subsection{Implementation of the SPINOR QUINT Model in \texttt{CLASS}}

To investigate the perturbative behavior of the SPINOR QUINT model, we incorporated the model into the public Einstein--Boltzmann code \texttt{CLASS}. The implementation was designed so that the background expansion generated by the spinor field is consistently propagated into the perturbation sector without modifying the Einstein--Boltzmann solver itself. The implementation consists of the following modifications.

\begin{itemize}
\item
The background module (\texttt{background.c}) was extended by introducing a new dark-energy equation-of-state identifier 
\texttt{SPINOR QUINT}, which evaluates the equation of state
\begin{equation}
w(a)=w_{\rm de},
\end{equation}
where $w_{\rm de}$ is constant throughout the cosmological evolution.

\item
Using the spinor-field solution obtained in Subsec.~2.1, the corresponding energy density evolves as
\begin{equation}
\rho_{\rm de}(a)=\rho_{{\rm de},0}a^{-3(1+w_{\rm de})},
\label{eq:rhoa}
\end{equation}
which is evaluated at every integration step of the background evolution.

\item
The input parser (\texttt{input.c}) was modified to allow the new equation of state through the keywords
\texttt{fluid\_equation\_of\_state = SPINOR QUINT}, together with the model parameter \texttt{w0\_fld}.

\item
The header file (\texttt{background.h}) was updated by adding the identifier \texttt{SPINOR QUINT} to the list of available dark-energy parameterizations.
\end{itemize}

In the present effective-fluid treatment, we adopt a rest-frame sound speed $c_s^2=1$, corresponding to the canonical quintessence-like perturbative prescription. This assumption is used to close the effective fluid perturbation system and should not be interpreted as a derivation of the spinor-field sound speed from first principles.
The perturbation equations remain identical to those already implemented in \texttt{CLASS}. Consequently, no modification of \texttt{perturbations.c} is required. Instead, the perturbation module automatically evolves the density and velocity perturbations using the background quantities supplied by the newly implemented equation of state. In the \texttt{CLASS}
calculation, the baryon physical density and the primordial
perturbation parameters are fixed to the Planck 2018--consistent
fiducial values $\omega_b=0.0224$, $A_s=2.1\times10^{-9}$,
$n_s=0.965$, and $\tau_{\rm reio}=0.054$ \cite{Planck2018}.
Thus, the MCMC analysis varies only the model parameters
$(w_{\rm de},\Omega_{m0},H_0)$, while the remaining early-Universe parameters are held fixed.

In addition to the background expansion history, the
\texttt{CLASS} implementation enables us to consistently follow the
linear perturbation evolution of the SPINOR QUINT field and its impact
on the matter and CMB observables. After these modifications, \texttt{CLASS} computes self-consistently the cosmological observables relevant for this work, including the expansion history, matter transfer functions, matter power spectrum, growth history, and the CMB temperature anisotropy spectrum.

\subsection{Cosmological Observables}

To confront the proposed model with cosmological observations, we consider several perturbation observables computed by \texttt{CLASS}.
The matter power spectrum is defined through the two-point correlation function of the Fourier-space matter density contrast~\cite{Peebles1980},
\begin{equation}
P(k)=\left<|\delta(k)|^2\right>,
\end{equation}
where $k$ denotes the comoving wave number. The matter power spectrum quantifies the amplitude of matter clustering and is one of the primary probes of structure formation~\cite{Eisenstein2005, Alam2017}.

The linear growth rate is defined as~\cite{Peebles1980, Linder2005}
\begin{equation}
f(a)=\frac{d\ln D}{d\ln a},
\end{equation}
where $D(a)$ is the linear growth factor. The quantity describes the rate at which matter perturbations grow during cosmic evolution.
Observationally, redshift-space distortion (RSD) measurements constrain the combination~\cite{Kaiser1987, Guzzo2008}
\begin{equation}
f\sigma_8(z)=f(z)\sigma_8(z),
\end{equation}
where $\sigma_8$ denotes the root-mean-square amplitude of matter fluctuations within spheres of radius $8\,h^{-1}\,\mathrm{Mpc}$.

The effective growth index is defined as~\cite{Wang1998, Linder2005}
\begin{equation}
\gamma(z)=\frac{\ln f(z)}{\ln \Omega_m(z)},
\end{equation}
where $\Omega_m(z)$ is the matter density parameter at redshift $z$. The growth index provides a convenient diagnostic for distinguishing different dark-energy and modified-gravity scenarios~\cite{Polarski2008}.

The temperature anisotropy spectrum is expressed as~\cite{Hu1997, Ma1995}
\begin{equation}
D_\ell^{TT}=\frac{\ell(\ell+1)}{2\pi}C_\ell^{TT},
\end{equation}
where $C_\ell^{TT}$ denotes the angular temperature power spectrum. The acoustic peak structure of $D_\ell^{TT}$ provides strong constraints on the background expansion and the evolution of cosmological perturbations~\cite{Planck2018}.

\section{Observational Data and Statistical Method}
\label{sec:obervation}
To constrain the cosmological parameters of the proposed SPINOR QUINT model, we perform a Markov Chain Monte Carlo (MCMC) analysis using several complementary cosmological observations that probe both the background expansion history and the growth of cosmic structures. Specifically, we consider the Pantheon+ Type Ia supernova compilation, Cosmic Chronometers (CC), DESI DR2 baryon acoustic oscillation (BAO) measurements, redshift-space distortion (RSD) data, and the Planck 2018 CMB distance priors. These datasets provide strong and complementary constraints on the expansion history, large-scale structure formation, and the geometry of the Universe.

In the following, the datasets used in this study are briefly explained.
\begin{itemize}

\item \textbf{Type Ia Supernovae: Pantheon+ (PP)}

We employ the Pantheon+ compilation \cite{PantheonPlus2022}, consisting of 1701 light-curve measurements corresponding to 1588 spectroscopically confirmed Type Ia supernovae in the redshift interval $0.01<z<2.26$. Following the Pantheon+ analysis, only supernovae with $z>0.01$ are considered in order to reduce the influence of peculiar velocities. The corresponding chi-square statistic is
\begin{equation}
\chi_{\rm PP}^{2}=
\Delta\boldsymbol{\mu}^{\,T}C^{-1}
\Delta\boldsymbol{\mu}-
\frac{\left(\mathbf{1}^{T}C^{-1}\Delta\boldsymbol{\mu}\right)^2}
{\mathbf{1}^{T}C^{-1}\mathbf{1}},
\end{equation}
where $\Delta\boldsymbol{\mu}=\boldsymbol{\mu}_{\rm obs}-\boldsymbol{\mu}_{\rm th}$, $C$ denotes the full statistical and systematic covariance matrix, and $\mathbf{1}$ is a vector of ones.

\item \textbf{Cosmic Chronometers (CC)}

To constrain the expansion history directly, we use 15 Cosmic Chronometer measurements of the Hubble parameter covering the redshift range
$0<z<2$
\cite{Stern2010,Moresco2012,Moresco2016,Zhang2014, Simon2005}.
The theoretical prediction is
\begin{equation}
\mathbf H_{\rm th}(z)=H_0E(z),
\end{equation}
and the residual vector defining as
\begin{equation}
{\Delta  \mathbf H}= {\mathbf H}_{\rm obs}-{\mathbf H}_{\rm th},
\end{equation}
where \({\mathbf H}_{\rm obs}\) denotes the observed Hubble parameters. The corresponding chi-square is given by
\begin{equation}
\chi^2_{\rm CC}
={\Delta \mathbf H}^{\,T}\mathbf{C}_{\rm CC}^{-1}{\Delta \mathbf H},
\end{equation}
where \(\mathbf{C}_{\rm CC}\) is the full \(15\times15\) covariance
matrix of the CC measurements.

\item \textbf{Baryon Acoustic Oscillations (BAO)}

We employ the latest DESI DR2 BAO measurements
\cite{DESI_DR2_BAO_2025,DESI_DR2_Lya_2025},
covering redshifts $z=0.295$, $0.510$, $0.706$, $0.934$, $1.321$, $1.484$,
and $2.33$. The measured observables include $D_V/r_d$, $D_M/r_d$, and
$D_H/r_d$, where $r_d$ denotes the sound horizon at the baryon drag epoch.
The BAO likelihood is written as
\begin{equation}
\chi^2_{\rm BAO}
=\boldsymbol{\Delta}_{\rm BAO}^{\,T}
\mathbf{C}_{\rm BAO}^{-1}
\boldsymbol{\Delta}_{\rm BAO},
\end{equation}
where \(\mathbf{C}_{\rm BAO}\) denotes the \(13\times13\) covariance matrix of the DESI DR2 BAO measurements. The residual vector $\boldsymbol{\Delta}_{\rm BAO}$ defined as 
\begin{equation}
\boldsymbol{\Delta}_{\rm BAO}=
\boldsymbol{\mathcal{O}}_{\rm obs}-
\boldsymbol{\mathcal{O}}_{\rm th},
\end{equation}
where $\boldsymbol{\mathcal{O}}_{\rm obs}$ and $\boldsymbol{\mathcal{O}}_{\rm th}$ are the observed and theoretical BAO measurements respectively.

\item \textbf{Redshift-Space Distortions (RSD)}

To constrain the growth of matter perturbations, we use 22 measurements of the growth observable
$f\sigma_8(z)$ compiled from various galaxy surveys over $0<z<2$ \cite{Nesseris2017}.
The corresponding chi-square statistic is
\begin{equation}
\chi_{\rm RSD}^{2}=\sum_i\frac{\left[f\sigma_{8,\rm obs}(z_i)
-f\sigma_{8,\rm th}(z_i)\right]^2}{\sigma_i^{2}}.
\end{equation}

\item \textbf{Planck 2018 CMB Distance Priors}

To incorporate constraints from the early Universe, we employ the compressed Planck 2018
TT, TE, EE + lowE distance priors \cite{Planck2018}.
The likelihood is constructed using the shift parameter $R$, the acoustic scale $\ell_A$, and the physical baryon density $\Omega_bh^2$ evaluated at the photon decoupling redshift
$z_*=1089.92$.

The corresponding chi-square is
\begin{equation}
\chi_{\rm CMB}^{2}=\Delta\mathbf{Y}^{T}
C_{\rm CMB}^{-1}\Delta\mathbf{Y},
\end{equation}
where
\begin{equation}
\Delta\mathbf{Y}=\left(R-R_{\rm obs},\;
\ell_A-\ell_{A,\rm obs},\;
\Omega_bh^2-\Omega_{b}h^2_{\rm obs}
\right).
\end{equation}

\end{itemize}

The cosmological parameters are estimated using the affine-invariant MCMC sampler implemented in the \texttt{emcee} package \cite{ForemanMackey2013} through the Python interface of the modified \texttt{CLASS} Boltzmann solver.
The primary analysis is performed using the combined
PP+CC+BAO+RSD+CMB dataset, while intermediate combinations
PP+CC+BAO and PP+CC+BAO+RSD are also considered to illustrate the impact of successive observational constraints. For comparison, the corresponding $\Lambda$CDM and $w$CDM cosmologies are also constrained using the full combined dataset and their respective best-fit parameters are used for perturbation studies.

Assuming independent datasets, the total likelihood is
\begin{equation}
\mathcal{L}_{\rm tot}
\propto
\exp
\left(
-\frac{\chi_{\rm tot}^{2}}{2}
\right),
\end{equation}
where
\begin{equation}
\chi_{\rm tot}^{2}
=\chi_{\rm PP}^{2}+\chi_{\rm CC}^{2}
+\chi_{\rm BAO}^{2}+\chi_{\rm RSD}^{2}
+\chi_{\rm CMB}^{2}.
\end{equation}
Uniform priors are adopted over the parameter ranges
\[
-1.2<w_{\rm de}<0.5,\qquad
0.01<\Omega_{m0}<0.5,\qquad
50<H_0<90.
\]

To compare the statistical performance of different cosmological models, we also compute the Akaike Information Criterion (AIC)~\cite{Akaike1974} and Bayesian Information Criterion (BIC)~\cite{Schwarz1978},
\begin{subequations}
\begin{align}
{\rm AIC}&=\chi_{\rm min}^{2}+2k,\\
{\rm BIC}&=\chi_{\rm min}^{2}+k\ln N,
\end{align}
\end{subequations}
where $k$ denotes the number of free model parameters and $N$ is the total number of observational data points.

\section{Results}
\label{sec:result}
\begin{figure}[!ht]
\centering
\includegraphics[width=0.8\textwidth]{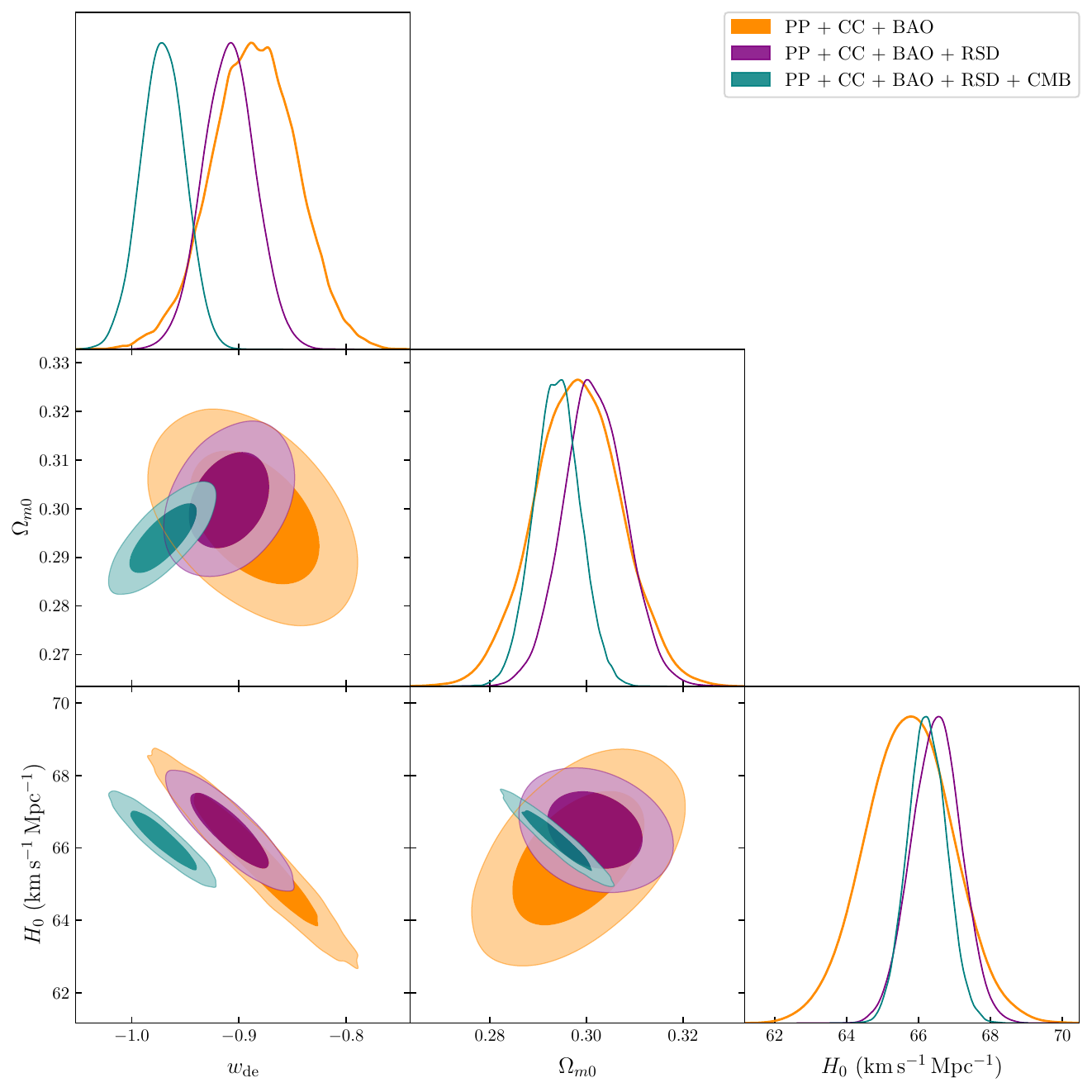}
\caption{One-dimensional marginalized posterior distributions and two-dimensional confidence contours (68\% and 95\% confidence levels) for the free parameters $(w_{\rm de},\,\Omega_{m0},\,H_0)$ of the SPINOR QUINT model obtained from the three observational dataset combinations.}
\label{fig:getdist}
\end{figure}

The free parameters of the SPINOR QUINT model are constrained using
the Markov Chain Monte Carlo (MCMC) technique implemented with the
\texttt{emcee} Python package and interfaced with the
Einstein--Boltzmann solver \texttt{CLASS}. The parameter estimation is
performed for three progressively combined observational datasets:
PP+CC+BAO(DESI DR2), PP+CC+BAO(DESI DR2)+RSD, and
PP+CC+BAO(DESI DR2)+RSD+CMB. The \texttt{CLASS} implementation
allows the background and linear perturbation evolution of the
SPINOR QUINT model to be calculated self-consistently, enabling us to
investigate both the expansion history and the growth of cosmic
structure. The marginalized posterior constraints are summarized in
Table~\ref{tab:spinor_constraints}, while the corresponding
one-dimensional marginalized posterior distributions and
two-dimensional confidence contours for the free parameters
$(w_{\rm de},\Omega_{m0},H_0)$ are shown in Fig.~\ref{fig:getdist}.

As progressively more observational probes are combined, the allowed
parameter space becomes increasingly restricted. For the
PP+CC+BAO combination, the marginalized constraints are
$w_{\rm de}=-0.88569^{+0.03897}_{-0.03895}$,
$\Omega_{m0}=0.29819^{+0.00882}_{-0.00868}$, and
$H_0=65.747^{+1.195}_{-1.207},
{\rm km\,s^{-1}\,Mpc^{-1}}$.
Upon including the RSD measurements, the constraints become
$w_{\rm de}=-0.90892^{+0.02406}_{-0.02450}$,
$\Omega_{m0}=0.30158^{+0.00653}_{-0.00612}$, and
$H_0=66.499^{+0.672}_{-0.695}\,
{\rm km\,s^{-1}\,Mpc^{-1}}$.
Thus, the inclusion of RSD data reduces the uncertainty in the
dark-energy equation-of-state parameter and provides an important
additional constraint through the growth of cosmic structure.

The addition of the CMB distance-prior information further narrows
the posterior distributions. For the full
PP+CC+BAO+RSD+CMB combination, we obtain
$w_{\rm de}=-0.97102^{+0.02061}_{-0.02061}$,
$\Omega_{m0}=0.29391^{+0.00469}_{-0.00472}$, and
$H_0=66.232^{+0.546}_{-0.532}\,
{\rm km\,s^{-1}\,Mpc^{-1}}$.
The corresponding derived parameters are
$t_0 = 14.242^{+0.015}_{-0.014}\ {\rm Gyr}$,
$r_d = 150.801^{+0.189}_{-0.197}\ {\rm Mpc}$,
$\sigma_8 = 0.75370^{+0.00760}_{-0.00746}$,
$S_8 = 0.74591^{+0.00667}_{-0.00648}$.
The substantially tighter constraints obtained for $\sigma_8$ and
$S_8$ when the RSD and CMB information are included demonstrate the
importance of combining expansion-history and structure-growth
observables. In particular, the full dataset constrains the present-day clustering amplitude at approximately the percent level. The resulting growth observables and matter power spectrum are examined below using
the linear perturbation predictions of \texttt{CLASS}.

\begin{table*}
\centering
\caption{Posterior constraints on the SPINOR QUINT model for different combinations of observational datasets. The quantities $t_0$ and $r_d$ are derived from the background evolution, while $\sigma_8$ and $S_8$ are obtained from the linear perturbation
evolution implemented in \texttt{CLASS}.}
\label{tab:spinor_constraints}
\begin{tabular}{lccc}
\hline
Parameter &
PP+CC+BAO &
PP+CC+BAO+RSD &
PP+CC+BAO+RSD+CMB
\\
\hline

$w_{\rm de}$ &
$-0.88569^{+0.03897}_{-0.03895}$ &
$-0.90892^{+0.02406}_{-0.02450}$ &
$-0.97102^{+0.02061}_{-0.02061}$ \\

$\Omega_{m0}$ &
$0.29819^{+0.00882}_{-0.00868}$ &
$0.30158^{+0.00653}_{-0.00612}$ &
$0.29391^{+0.00469}_{-0.00472}$ \\

$H_0$ &
$65.747^{+1.195}_{-1.207}$ &
$66.499^{+0.672}_{-0.695}$ &
$66.232^{+0.546}_{-0.532}$ \\

$t_0$ [Gyr] &
$14.077^{+0.216}_{-0.220}$ &
$13.947^{+0.108}_{-0.106}$ &
$14.242^{+0.015}_{-0.014}$ \\

$r_d$ [Mpc] &
$150.725^{+1.903}_{-1.939}$ &
$149.580^{+0.896}_{-0.880}$ &
$150.801^{+0.189}_{-0.197}$ \\

$\sigma_8$ &
$0.73449^{+0.04130}_{-0.04075}$ &
$0.76000^{+0.01811}_{-0.01835}$ &
$0.75370^{+0.00760}_{-0.00746}$ \\

$S_8$ &
$0.73221^{+0.05095}_{-0.04879}$ &
$0.76207^{+0.02300}_{-0.02296}$ &
$0.74591^{+0.00667}_{-0.00648}$ \\

\hline
\end{tabular}
\end{table*}

Table~\ref{tab:spinor_comparison} presents a comparison of the
marginalized posterior constraints on the cosmological and derived
parameters for $\Lambda$CDM, $w$CDM, and the SPINOR QUINT model using
the full PP+CC+BAO+RSD+CMB dataset. All three models are
analyzed within the same MCMC framework and with the same
observational combination, providing a consistent basis for comparing
their inferred cosmological parameters and perturbation observables.

The constraints obtained for SPINOR QUINT are very close to those of
the phenomenological $w$CDM model. For SPINOR QUINT, we find
$w_{\rm de}=-0.9710^{+0.0206}_{-0.0206}$,
$\Omega_{m0}=0.2939^{+0.0047}_{-0.0047}$, and
$H_0=66.23^{+0.55}_{-0.53}\,
{\rm km\,s^{-1}\,Mpc^{-1}}$.
The corresponding $w$CDM constraints are
$w_{\rm de}=-0.9708^{+0.0202}_{-0.0204}$,
$\Omega_{m0}=0.2938^{+0.0047}_{-0.0046}$, and
$H_0=66.23^{+0.54}_{-0.53}\,
{\rm km\,s^{-1}\,Mpc^{-1}}$.
Thus, the posterior constraints on $\Omega_{m0}$ and $H_0$ are
essentially identical for the two dynamical dark-energy models, while
their inferred values of $w$ differ by less than $0.001$. The SPINOR
QUINT constraint is also consistent with the cosmological-constant
value $w_{\rm de}=-1$ at approximately the $1.4\sigma$ level. These results
indicate that, for the present dataset, the background expansion
history of the SPINOR QUINT model closely tracks that of a constant-$w$
dark-energy model.

For comparison, $\Lambda$CDM favors a somewhat lower matter density
and a higher Hubble constant,
$\Omega_{m0}=0.2888^{+0.0031}_{-0.0031}$,
$H_0=66.94^{+0.23}_{-0.22}\,
{\rm km\,s^{-1}\,Mpc^{-1}}$.
Despite these differences, the inferred cosmic ages are remarkably
similar:
$t_0=14.226^{+0.009}_{-0.009}\ {\rm Gyr}
(\Lambda{\rm CDM})$,
$t_0=14.243^{+0.007}_{-0.007}\ {\rm Gyr}
(w{\rm CDM})$,
$t_0=14.242^{+0.015}_{-0.014}\ {\rm Gyr}
({\rm SPINOR\ QUINT})$.
The similarity of the ages results from the compensating dependence of
the cosmic age on the inferred expansion history and the value of
$H_0$. Likewise, the sound horizon at the drag epoch is very similar
among the three models:
$r_d =150.655^{+0.159}_{-0.158}\ {\rm Mpc}
(\Lambda{\rm CDM})$,
$r_d =150.804^{+0.202}_{-0.199}\ {\rm Mpc}
(w{\rm CDM})$,
$r_d =150.801^{+0.189}_{-0.197}\ {\rm Mpc}
({\rm SPINOR\ QUINT})$.
This close agreement is expected because the three models employ the
same early-time baryonic and cold-dark-matter sectors, while their
principal differences arise from the late-time dark-energy dynamics.

The comparison becomes particularly relevant for the growth of
structure. For $\Lambda$CDM, we obtain
$\sigma_8=0.7637^{+0.0021}_{-0.0021}$ and
$S_8=0.7494^{+0.0060}_{-0.0061}$, whereas $w$CDM gives
$\sigma_8=0.7536^{+0.0061}_{-0.0061}$ and
$S_8=0.7458^{+0.0067}_{-0.0069}$. The corresponding SPINOR QUINT
constraints are
$\sigma_8=0.7537^{+0.0076}_{-0.0075}$ and
$S_8=0.7459^{+0.0067}_{-0.0065}$.
Thus, the predicted clustering amplitude and $S_8$ parameter for
SPINOR QUINT are essentially identical to those obtained in $w$CDM
for the full dataset. Both dynamical dark-energy models prefer a
slightly lower $\sigma_8$ than $\Lambda$CDM, although the differences
are small compared with the corresponding posterior uncertainties.
The agreement between SPINOR QUINT and $w$CDM in these derived
quantities provides an important consistency check of the perturbation
implementation.

\begin{table*}
\centering
\caption{Marginalized posterior constraints on the cosmological
parameters and derived quantities for $\Lambda$CDM, $w$CDM, and the
SPINOR QUINT model  using the combined
PP+CC+BAO(DESI DR2)+RSD+CMB dataset.}
\label{tab:spinor_comparison}
\begin{tabular}{lccc}
\hline
Parameter &
$\Lambda$CDM &
$w$CDM &
SPINOR QUINT \\
\hline
$w_{\rm de}$ &
$-1 (\rm fixed)$ &
$-0.9708^{+0.0202}_{-0.0204}$ &
$-0.9710^{+0.0206}_{-0.0206}$ \\

$\Omega_{m0}$ &
$0.2888^{+0.0031}_{-0.0031}$ &
$0.2938^{+0.0047}_{-0.0046}$ &
$0.2939^{+0.0047}_{-0.0047}$ \\

$H_0$ &
$66.94^{+0.23}_{-0.22}$ &
$66.23^{+0.54}_{-0.53}$ &
$66.23^{+0.55}_{-0.53}$ \\

$t_0\,[\mathrm{Gyr}]$ &
$14.226^{+0.009}_{-0.009}$ &
$14.243^{+0.007}_{-0.007}$ &
$14.242^{+0.015}_{-0.014}$ \\

$r_d\,[\mathrm{Mpc}]$ &
$150.655^{+0.159}_{-0.158}$ &
$150.804^{+0.202}_{-0.199}$ &
$150.801^{+0.189}_{-0.197}$ \\

$\sigma_8$ &
$0.7637^{+0.0021}_{-0.0021}$ &
$0.7536^{+0.0061}_{-0.0061}$ &
$0.7537^{+0.0076}_{-0.0075}$ \\

$S_8$ &
$0.7494^{+0.0060}_{-0.0061}$ &
$0.7458^{+0.0067}_{-0.0069}$ &
$0.7459^{+0.0067}_{-0.0065}$ \\
\hline
\end{tabular}
\label{tab:derived_constraints}
\end{table*}

The statistical comparison of the three models is summarized in
Table~\ref{tab:statistical_comparison}. For the combined
PP+CC+BAO(DESI DR2)+RSD+CMB dataset, the total $\chi^2$ values are
$1519.80$, $1517.80$, and $1517.78$ for $\Lambda$CDM, $w$CDM, and
SPINOR QUINT, respectively. The SPINOR QUINT model therefore provides
the lowest total $\chi^2$, but the improvement relative to $w$CDM is
only $\Delta\chi^2 \simeq 0.02$, which is statistically negligible. Relative to $\Lambda$CDM, the improvement is $\Delta\chi^2\simeq2.03$.
The corresponding information criteria are
$\mathrm{AIC}=1523.80$, $1523.80$, and $1523.78$ for $\Lambda$CDM,
$w$CDM, and SPINOR QUINT, respectively. The AIC therefore shows no
meaningful preference among the three models. The BIC values are
$1534.60$, $1540.00$, and $1539.98$, respectively. The higher BIC
values of the two dynamical dark-energy models relative to
$\Lambda$CDM reflect the additional free parameter $w_{\rm de}$ and the
associated complexity penalty. Consequently, although SPINOR QUINT
achieves a slightly lower $\chi^2$ than both $\Lambda$CDM and $w$CDM,
the statistical improvement is insufficient to provide evidence for
the additional degree of freedom.

Overall, the information-criterion analysis indicates that the current
combined dataset does not statistically distinguish SPINOR QUINT from
the simpler $\Lambda$CDM or phenomenological $w$CDM descriptions.
Nevertheless, the fact that SPINOR QUINT reproduces the observational
constraints and the perturbation observables of $w$CDM with a
comparable goodness of fit is an important result, since the
constant-$w$ behavior arises from the underlying spinor-field
realization rather than being introduced purely phenomenologically.

\begin{table*}
\centering
\caption{Statistical comparison of the $\Lambda$CDM, $w$CDM, and
SPINOR QUINT (SQ) models using the combined
PP+CC+BAO(DESI DR2)+RSD+CMB dataset.}
\label{tab:statistical_comparison}
\begin{tabular}{lccc}
\hline
 & $\Lambda$CDM & $w$CDM & SQ \\
\hline
$\chi^2_{\rm PP}$  & 1445.20 & 1439.48 & 1439.49 \\
$\chi^2_{\rm CC}$  & 6.69    & 6.67    & 6.67    \\
$\chi^2_{\rm BAO}$ & 35.19   & 33.57   & 33.71   \\
$\chi^2_{\rm RSD}$ & 12.29   & 12.68   & 12.66   \\
$\chi^2_{\rm CMB}$ & 20.44   & 25.41   & 25.25   \\
\hline
$\chi^2_{\rm tot}$ & 1519.80 & 1517.80 & 1517.78 \\
$\chi^2_{\rm red}$ & 0.9284  & 0.9278  & 0.9277  \\
AIC                & 1523.80 & 1523.80 & 1523.78 \\
BIC                & 1534.60 & 1540.00 & 1539.98 \\
\hline
\end{tabular}
\end{table*}

We next examine the linear perturbation predictions of the three
models in order to assess whether the spinor-field realization leads
to observable differences in the growth of cosmic structure and in the
CMB temperature anisotropy spectrum.

\begin{figure}[!ht]
\centering
\includegraphics[width=0.8\textwidth]{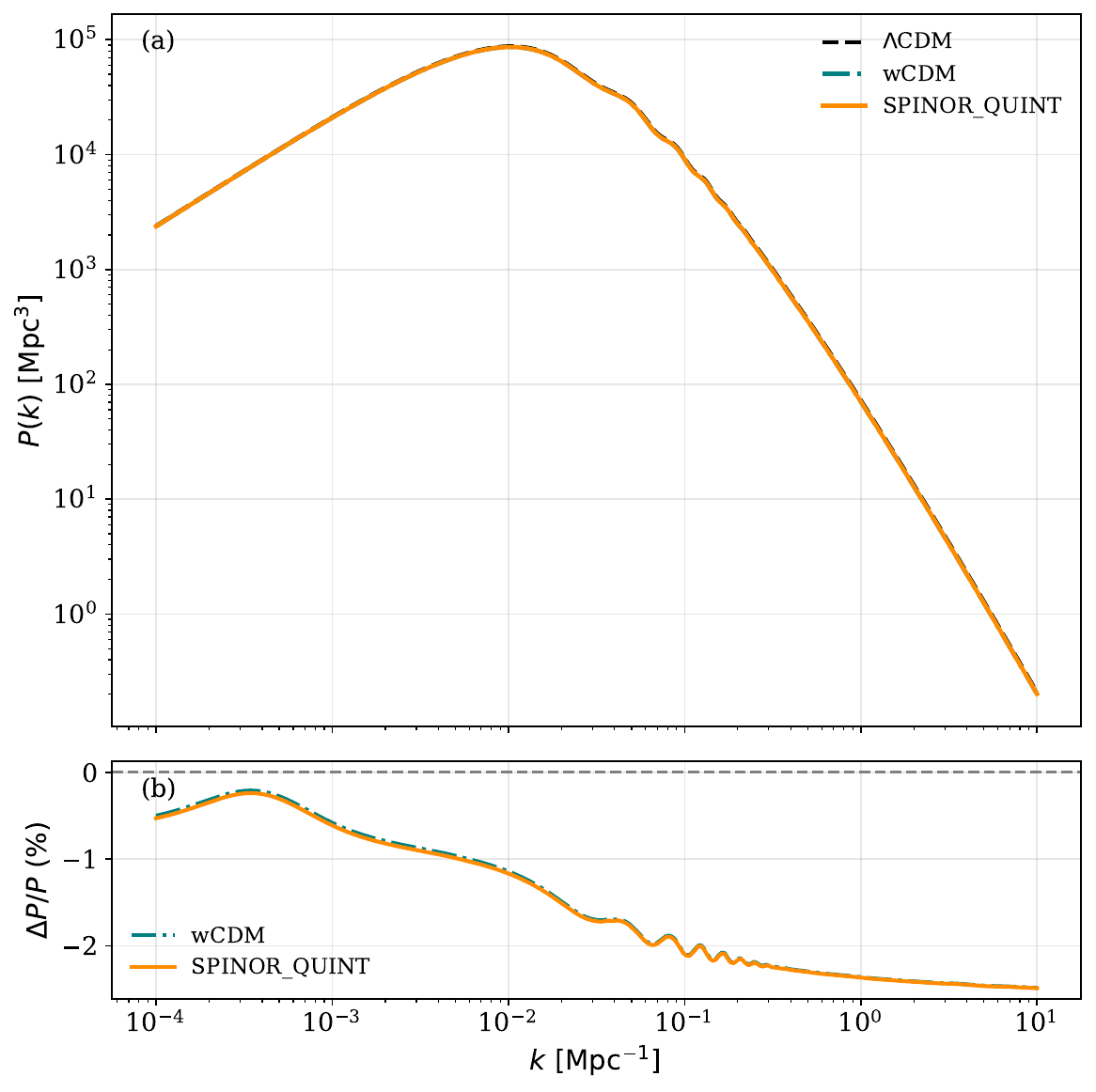}
\caption{
Present-day linear matter power spectrum for the best-fit
$\Lambda$CDM, $w$CDM, and SPINOR QUINT models.
Upper panel (a): matter power spectrum $P(k)$ at $z=0$.
Lower panel (b): fractional difference,
$\Delta P/P=\left(P_{\rm model}-P_{\Lambda{\rm CDM}}\right)/P_{\Lambda{\rm CDM}}$,
for the $w$CDM and SPINOR QUINT models relative to the $\Lambda$CDM
prediction.
}
\label{fig:pk}
\end{figure}

Figure~\ref{fig:pk} shows the present-day linear matter power spectrum
$P(k)$ predicted by the $\Lambda$CDM, $w$CDM, and SPINOR QUINT models,
together with their fractional differences relative to $\Lambda$CDM.
The three spectra are very similar over the range of wavenumbers
considered. The corresponding fractional residuals show that the
predictions of both $w$CDM and SPINOR QUINT remain at the percent level relative to $\Lambda$CDM. In particular, the SPINOR QUINT prediction
closely follows the $w$CDM result, consistent with the similarity of
their posterior constraints on $w_{\rm de}$, $\Omega_{m0}$, and $H_0$. Thus,
within the range of scales explored here, the spinor-field realization
does not introduce a substantial modification to the linear matter
power spectrum relative to the corresponding constant-$w$ model.

\begin{figure}[!ht]
\centering
\includegraphics[width=0.8\textwidth]{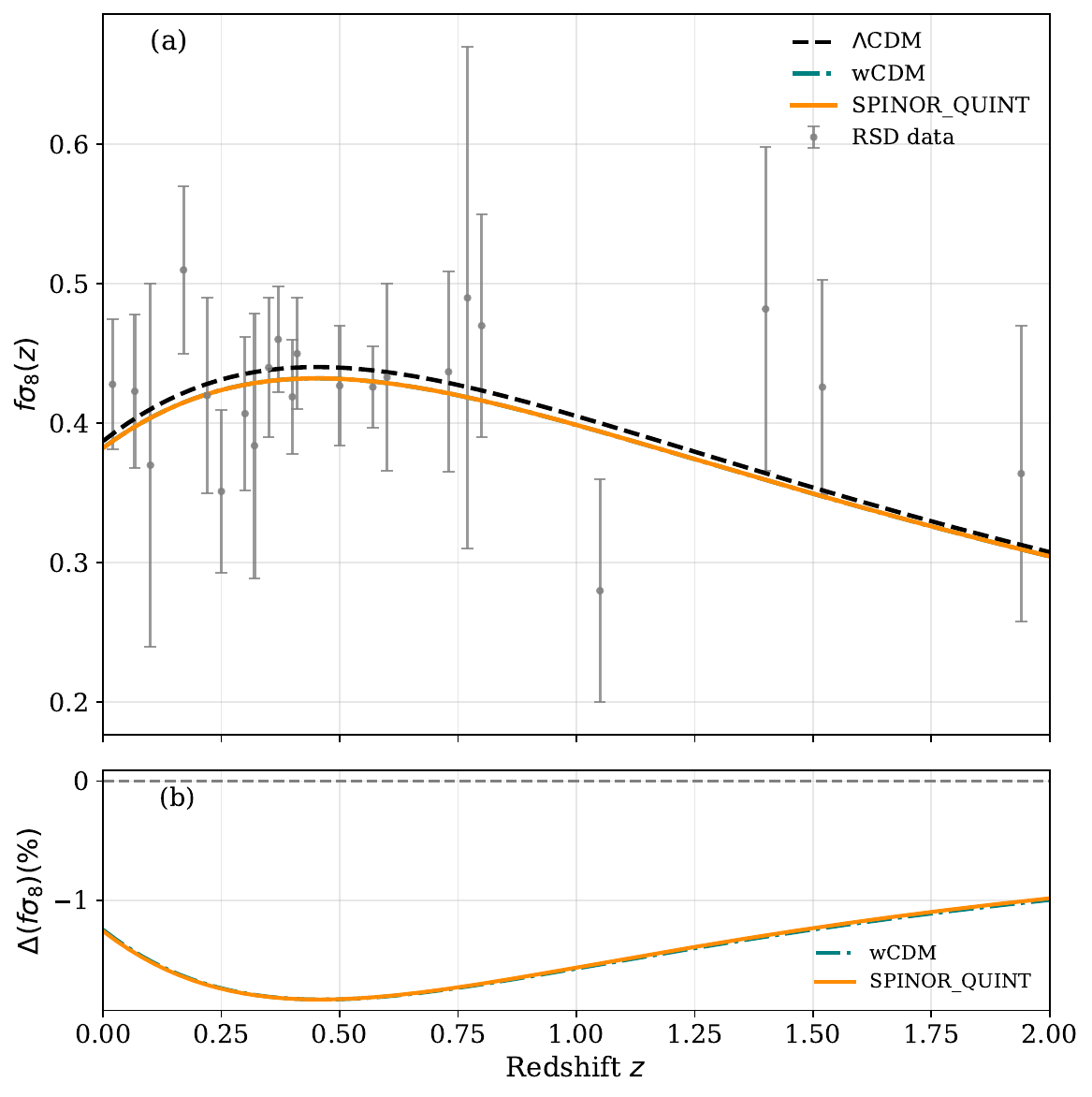}
\caption{
Evolution of the growth-rate observable $f\sigma_8(z)$ for the best-fit
$\Lambda$CDM, $w$CDM, and SPINOR QUINT models.
Upper panel (a): comparison of the theoretical predictions with the
RSD measurements \cite{Nesseris2017}.
Lower panel (b): fractional difference,
$\Delta(f\sigma_8)=100\times[(f\sigma_8)_{\rm model}-(f\sigma_8)_{\Lambda{\rm CDM}}]/(f\sigma_8)_{\Lambda{\rm CDM}}$,
for the $w$CDM and SPINOR QUINT models relative to the $\Lambda$CDM prediction.
}
\label{fig:fsigma8}
\end{figure}

This behavior is further illustrated by the growth-rate observable
$f\sigma_8(z)$. Figure~\ref{fig:fsigma8} shows the predictions of the
$\Lambda$CDM, $w$CDM, and SPINOR QUINT models together with the
available RSD measurements \cite{Nesseris2017}. The three theoretical
predictions follow the observed growth history over the redshift range
considered and remain close to one another throughout the interval.
The lower panel shows the fractional deviation from the $\Lambda$CDM
prediction. Both $w$CDM and SPINOR QUINT exhibit deviations of less
than approximately $1\%$ over the investigated redshift range, with
the largest differences occurring at intermediate redshifts. The close
agreement between SPINOR QUINT and $w$CDM is consistent with their
nearly identical constraints on $\sigma_8$ and $S_8$ obtained from the
full dataset.

\begin{figure}[!ht]
\centering
\includegraphics[width=0.8\textwidth]{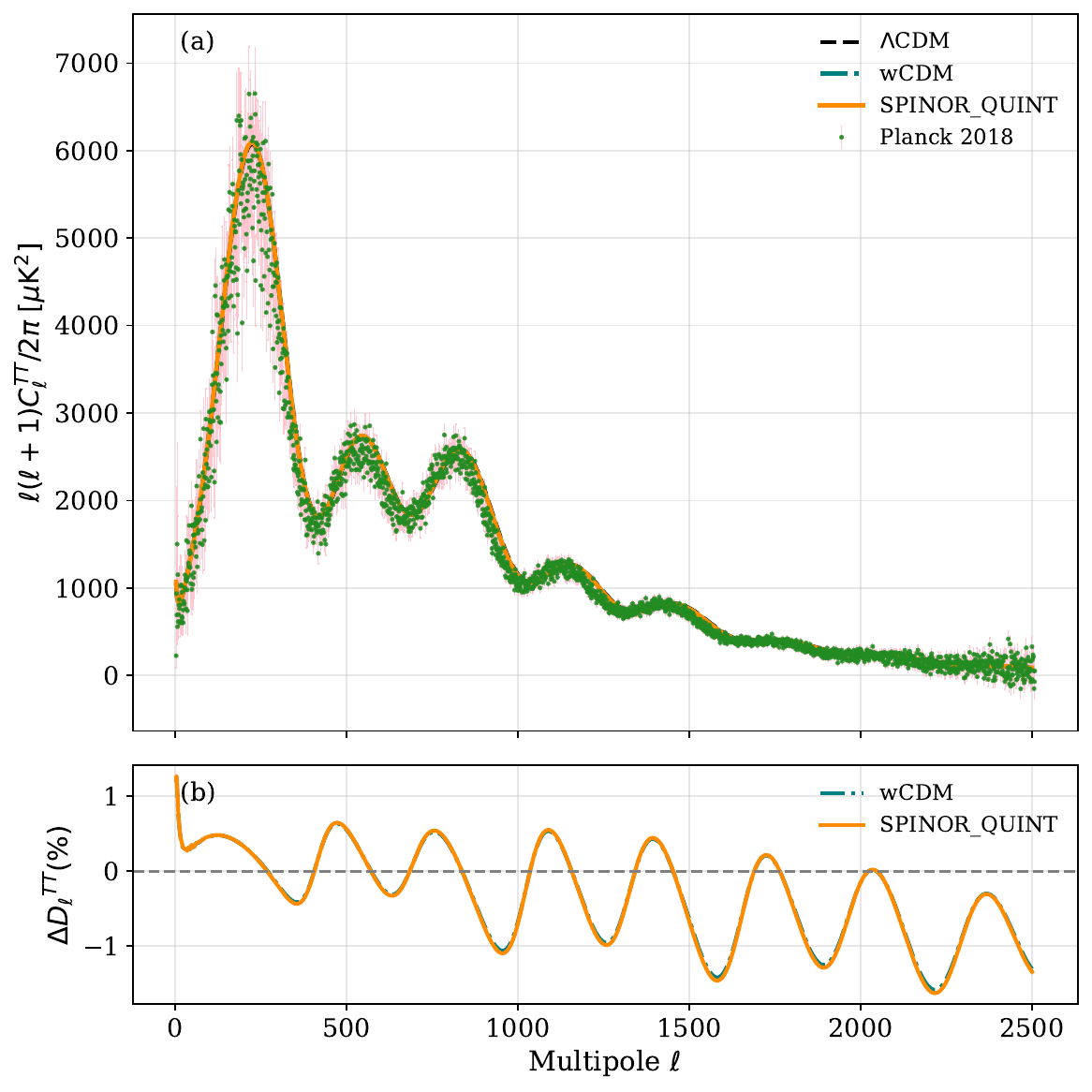}
\caption{
CMB temperature anisotropy power spectrum for the best-fit
$\Lambda$CDM, $w$CDM, and SPINOR QUINT models compared with the
\textit{Planck} 2018 TT measurements.
Upper panel (a): temperature power spectrum,
$D_\ell^{TT}=\ell(\ell+1)C_\ell^{TT}/2\pi$.
Lower panel (b): fractional difference with respect to the
$\Lambda$CDM prediction,
$100\times(D_\ell^{\rm model}-D_\ell^{\Lambda{\rm CDM}})/D_\ell^{\Lambda{\rm CDM}}$.
The deviations remain below approximately $0.5\%$ over the full multipole range, indicating that the SPINOR QUINT model reproduces the observed CMB temperature anisotropies with an accuracy comparable to that of the standard cosmological model.
}
\label{fig:CMB_TT}
\end{figure}

The CMB temperature anisotropy spectra predicted by the
$\Lambda$CDM, $w$CDM, and SPINOR QUINT models are shown in
Fig.~\ref{fig:CMB_TT}, together with the \textit{Planck} 2018 TT
measurements. These spectra are calculated independently with
\texttt{CLASS} using the cosmological parameter constraints obtained
from the combined likelihood, which employs the CMB distance-prior
information. The three models reproduce the overall acoustic-peak
structure in a very similar manner. In particular, the positions of
the principal acoustic features are nearly coincident, reflecting the
similarity of the inferred distance to last scattering and the
early-time cosmological parameters.

The lower panel shows the fractional difference with respect to the
$\Lambda$CDM prediction,
\begin{equation}
100\times
\frac{D_{\ell}^{TT,{\rm model}}-D_{\ell}^{TT,\Lambda{\rm CDM}}}
{D_{\ell}^{TT,\Lambda{\rm CDM}}}.
\end{equation}

The predicted spectra of $w$CDM and SPINOR QUINT remain very close to
the $\Lambda$CDM spectrum over most of the multipole range, with only
small differences at low multipoles. This agreement provides an
additional consistency check of the perturbation implementation.
Importantly, the plotted full TT spectra should not be interpreted as
a replacement for a full CMB power-spectrum likelihood, since the
parameter constraints presented here use compressed CMB distance
priors.

\begin{figure}[!ht]
\centering
\includegraphics[width=0.8\textwidth]{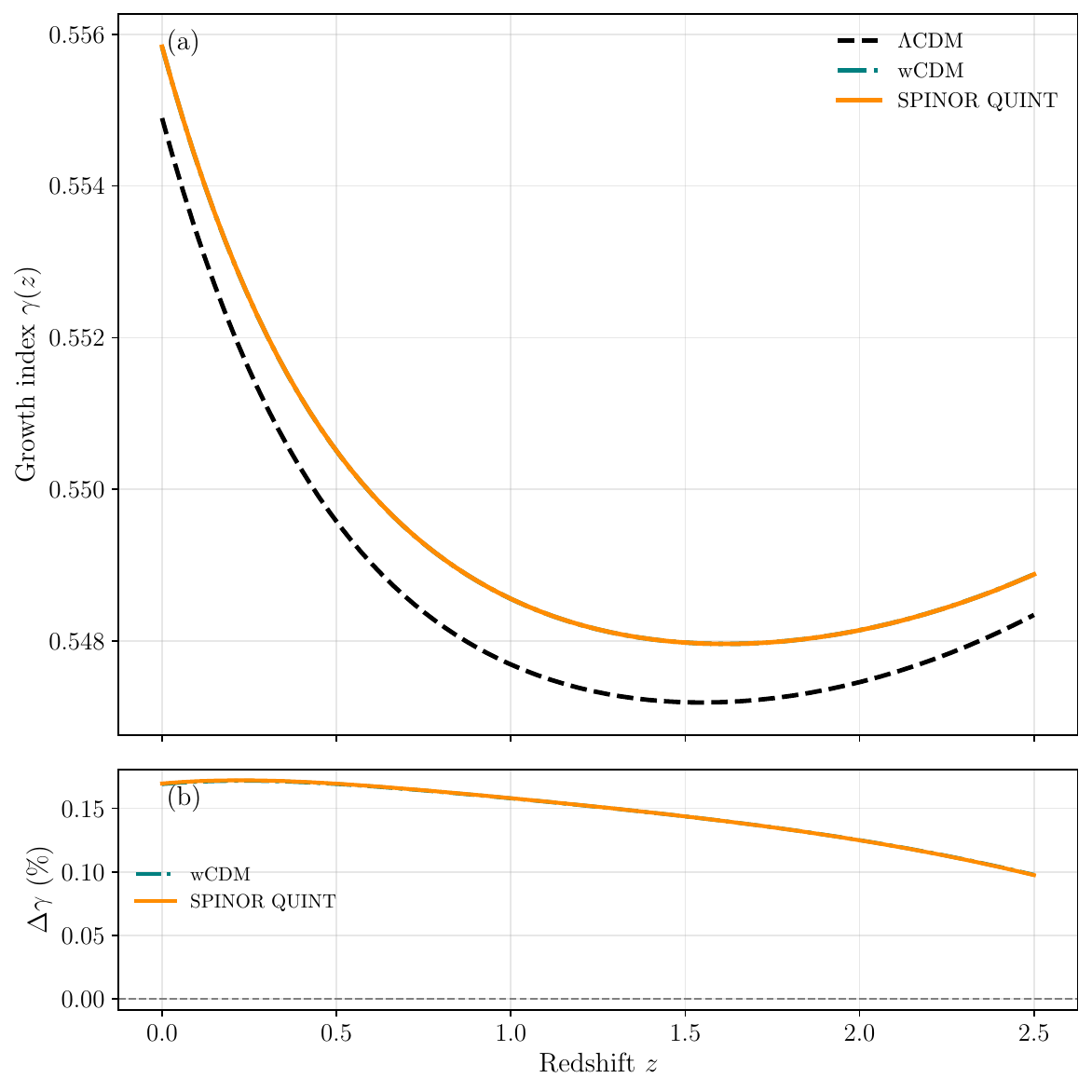}
\caption{
Growth index for the best-fit $\Lambda$CDM, $w$CDM, and SPINOR QUINT models.
Upper panel (a): evolution of the growth index $\gamma(z)$.
Lower panel (b): fractional difference,
$100\times(\gamma_{\rm model}-\gamma_{\Lambda{\rm CDM}})/\gamma_{\Lambda{\rm CDM}}$,
relative to the $\Lambda$CDM prediction. The deviations remain below approximately $0.1\%$ over the entire redshift range, demonstrating that the SPINOR QUINT model closely follows the standard growth history of cosmic structures.
}
\label{fig:growth_gamma}
\end{figure}

The growth index $\gamma(z)$ provides another diagnostic of the
response of matter perturbations to the underlying expansion history.
Figure~\ref{fig:growth_gamma} shows the evolution of $\gamma(z)$ for
the $\Lambda$CDM, $w$CDM, and SPINOR QUINT models over
$0\leq z\leq2.5$. The three models exhibit very similar evolution,
with $\gamma$ remaining close to the standard $\Lambda$CDM-like
behavior throughout the redshift interval. At $z=0$, the predicted
values are approximately $\gamma\simeq0.555$, while they approach
$\gamma\simeq0.547$ at $z\simeq2.5$.

The lower panel displays the fractional difference relative to
$\Lambda$CDM,
\begin{equation}
100\times
\frac{\gamma_{\rm model}-\gamma_{\Lambda{\rm CDM}}}
{\gamma_{\Lambda{\rm CDM}}}.
\end{equation}
The deviations remain below approximately $0.1\%$ over the redshift
range considered. The close agreement between SPINOR QUINT and
$w$CDM indicates that, for the parameter region favored by the full
dataset, the spinor-field realization produces a growth history very
similar to that of a constant-$w$ dark-energy model.

\section{Discussion and Conclusions}
\label{sec:conclusion}
In this work, we have investigated the cosmological implications of a
spinor-field realization of quintessence beyond the homogeneous
background level by implementing the model in the
\texttt{CLASS} Einstein--Boltzmann solver. The free parameters of the
SPINOR QUINT model were constrained using a Markov Chain Monte Carlo
analysis with the \texttt{emcee} package, employing the combined
Pantheon$^{+}$ Type-Ia supernovae, cosmic chronometer (CC), DESI DR2
BAO, redshift-space distortion (RSD), and CMB distance-prior
observations. In contrast to a purely background-level analysis, the
modified \texttt{CLASS} implementation allows the evolution of linear
perturbations to be followed consistently and provides predictions for
the matter clustering and CMB observables.

We considered three progressively combined data sets,
PP+CC+BAO, PP+CC+BAO+RSD, and
PP+CC+BAO+RSD+CMB. The addition of RSD data substantially improves
the constraints on the equation-of-state parameter and the matter
density, while the inclusion of the compressed CMB distance priors
provides the tightest constraints. For the full
PP+CC+BAO+RSD+CMB combination, we obtain
$w_{\rm de}=-0.9710^{+0.0206}_{-0.0206}$,
$\Omega_{m0}=0.2939^{+0.0047}_{-0.0047}$, and
$H_0=66.23^{+0.55}_{-0.53}\,
\mathrm{km\,s^{-1}\,Mpc^{-1}}$. The corresponding derived quantities
are $t_0=14.242^{+0.015}_{-0.014}$ Gyr,
$r_d=150.801^{+0.189}_{-0.197}$ Mpc,
$\sigma_8=0.7537^{+0.0076}_{-0.0075}$, and
$S_8=0.7459^{+0.0067}_{-0.0065}$.
Thus, the preferred equation of state is close to the cosmological
constant value, while the inferred clustering amplitude is somewhat
lower than the value commonly obtained from the Planck 2018
$\Lambda$CDM analysis.

Although the relatively low value of $S_8$ is potentially interesting
in the context of the weak-lensing clustering-amplitude discrepancy,
our results should not be interpreted as a resolution of the
$S_8$ tension. The present analysis does not include a dedicated
weak-lensing likelihood and the difference in $S_8$ is not sufficient
by itself to establish a statistically significant alleviation of the
tension. Rather, the main result is that the spinor-field realization
remains compatible with the observed amplitude of matter clustering
when the model is tested at the linear perturbation level. This
provides an important consistency check of the spinor-quintessence
scenario beyond its background expansion history.

Since the background evolution of the present SPINOR QUINT
realization is mathematically equivalent to that of a constant-$w$
dark-energy model, we performed a direct comparison with
phenomenological $w$CDM and standard $\Lambda$CDM using the same
likelihoods, data sets, and MCMC methodology. For the full
PP+CC+BAO+RSD+CMB combination, the SPINOR QUINT and $w$CDM models
yield remarkably similar constraints. In particular, both give
$w_{\rm de}\simeq-0.971$, $\Omega_{m0}\simeq0.294$, and
$H_0\simeq66.23~\mathrm{km\,s^{-1}\,Mpc^{-1}}$, with nearly identical
predictions for $t_0$, $r_d$, $\sigma_8$, and $S_8$. This agreement is
expected at the background level from their common constant-$w$
expansion history, but the CLASS implementation allows this
correspondence to be tested consistently at the perturbation level.

The statistical comparison further shows that the three models provide
essentially indistinguishable fits to the combined observational data.
The total chi-square values are
$\chi^2_{\rm tot}=1519.80$, $1517.80$, and $1517.78$ for
$\Lambda$CDM, $w$CDM, and SPINOR QUINT, respectively. Although
SPINOR QUINT gives the smallest total $\chi^2$, its improvement over
$w$CDM is only $\Delta\chi^2\simeq0.02$, which is statistically
negligible. The AIC values are likewise almost identical,
whereas the BIC favors $\Lambda$CDM because of its smaller number of
free parameters. Therefore, the present data do not provide
statistically significant evidence in favor of the additional
dynamical degree of freedom represented by the SPINOR QUINT model.
Nevertheless, the absence of a significant statistical penalty in the
likelihood and the close agreement with the perturbation observables
demonstrate that the model remains a viable alternative realization of
late-time dark-energy phenomenology.

We have also examined several observables that directly probe the
linear perturbation sector. The present-day matter power spectrum
$P(k)$, the growth-rate observable $f\sigma_8(z)$, the CMB temperature
anisotropy spectrum $D_\ell^{TT}$, and the growth index $\gamma(z)$
predicted by SPINOR QUINT closely follow the corresponding
$\Lambda$CDM and $w$CDM predictions. The deviations from $\Lambda$CDM
remain at the sub-percent level over the scales and redshift ranges
considered, with the largest differences occurring in the growth and
low-multipole CMB observables. In particular, the nearly identical
$P(k)$ and $f\sigma_8(z)$ predictions demonstrate that the spinor-field
realization does not introduce significant departures from the
standard growth history within the parameter region favored by the
current data.

An important aspect of this analysis is the extension of the standard
\texttt{CLASS} framework to accommodate the SPINOR QUINT realization.
We introduced the corresponding ``SPINOR\_QUINT'' fluid type into the
\texttt{background.c} and \texttt{input.c} modules of \texttt{CLASS}, together with the
appropriate equation-of-state evolution, while adopting the canonical
effective sound speed $c_s^2=1$. This implementation provides a
consistent framework for computing both the background evolution and
linear perturbations of the model and therefore enables direct
comparison with observables sensitive to cosmic structure formation. The spinor model provides a field-theoretic realization of constant-$w$ dark energy, and when its effective perturbative sound speed is chosen as $c_s^2=1$, its linear cosmological predictions coincide closely with those of phenomenological $w$CDM.

The results presented here establish the consistency of the
spinor-quintessence scenario at the linear perturbation level, but they
also indicate that the present observational combination does not
strongly distinguish it from phenomenological constant-$w$ dark energy.
A more stringent test will require additional observables that are
directly sensitive to the growth of structure. In particular, future
analyses incorporating weak-lensing measurements, galaxy clustering,
full large-scale-structure likelihoods, and the full CMB temperature
and polarization spectra will provide more powerful tests of the
perturbation sector. Such extensions will allow us to determine whether
the underlying spinor-field dynamics can produce observable signatures
that distinguish SPINOR QUINT from both $w$CDM and $\Lambda$CDM.

{\bf Declarations}
\vskip 3 mm {\bf Competing interests:} { There is
no conflict of interests.}
\vskip 3 mm {\bf CRediT author statement:} { \textbf{M.Goray} is a sole author.
\vskip 3 mm {\bf Funding:}  Not applicable.
\vskip 3 mm {\bf Availability of data and materials:} No new data sets were generated during the current study.

\appendix
\section*{Appendix A: Implementation of the SPINOR QUINT Model in CLASS}
\label{app:class_implementation}

To compute the linear perturbation observables of the SPINOR QUINT model, we implemented the model as an additional dark-energy fluid option in the public Einstein--Boltzmann solver \texttt{CLASS}. The modification is restricted to the background-fluid equation-of-state interface, while the standard perturbation and Boltzmann evolution routines of \texttt{CLASS} are retained. The implementation therefore allows the spinor-quintessence background to be propagated consistently into the linear perturbation calculation.

First, a new equation-of-state identifier, \texttt{SPINOR\_QUINT}, was introduced in the equation-of-state enumeration in \texttt{include/background.h}:

\begin{verbatim}
enum equation_of_state {CLP,EDE,SPINOR_QUINT};
\end{verbatim}

The corresponding input option was then added to \texttt{source/input.c}, allowing the model to be selected through the standard \texttt{fluid\_equation\_of\_state} CLASS input:

\begin{verbatim}
else if ((strstr(string1,"SPINOR_QUINT") != NULL) ||
(strstr(string1,"spinor_quint") != NULL)) {
pba->fluid_equation_of_state = SPINOR_QUINT;
}
\end{verbatim}
\noindent
For this model, the equation-of-state parameter and the effective sound speed are read through the standard CLASS fluid parameters:

\begin{verbatim}
if (pba->fluid_equation_of_state == SPINOR_QUINT) {
class_read_double("w0_fld",pba->w0_fld);
class_read_double("cs2_fld",pba->cs2_fld);
}
\end{verbatim}

The background evolution is implemented in \texttt{source/background.c}. For the SPINOR QUINT model, the equation of state is constant,
$w_{\rm de}(a)=w_0$, and hence $\frac{dw_{\rm de}}{da}=0$. The corresponding CLASS implementation is

\begin{verbatim}
case SPINOR_QUINT:
*w_fld = pba->w0_fld;
break;
\end{verbatim}
\noindent
and
\begin{verbatim}
case SPINOR_QUINT:
*dw_over_da_fld = 0.;
break;
\end{verbatim}
\noindent
The analytic integral required by the CLASS background initial conditions is implemented as
\begin{verbatim}
case SPINOR_QUINT:
*integral_fld =
3.*(1.+pba->w0_fld)*log(1./a);
break;
\end{verbatim}
\noindent
which corresponds to the SPINOR QUINT EoS density evolution in Eq.~\eqref{mchapsped}.

The modified \texttt{CLASS} implementation used for the perturbation-level analysis, including the \texttt{SPINOR\_QUINT} fluid implementation, is publicly available at
\url{https://github.com/gorayphy/CLASS-SPINOR-QUINT#class-spinor-quint}.

\normalfont
\end{document}